# Superconductivity in the WP single crystal

Ziyi Liu[1,2†], Wei Wu[1†*], Zhenzheng Zhao[1], Hengcan Zhao[1], Jian Cui[1], Pengfei Shan[2], Jiahao Zhang[1], Changli Yang[1,3], Peijie Sun[1], Yu Sui[2], Jinguang Cheng[1,4], Li Lu[1,3], Jianlin Luo[1,3*], and Guangtong Liu[1,4*]

[1]Beijing National Laboratory of Condensed Matter Physics and Institute of Physics, Chinese Academy of Sciences, Beijing 100190, China

[2]School of Physics, Harbin Institute of Technology, Harbin, Heilongjiang 150001, China

[3] Collaborative Innovation Center of Quantum Matter, Beijing 100871, China

[4]Songshan Lake Materials Laboratory, Dongguan, Guangdong 523808, China

† These authors contributed equally to this work. Correspondence and requests for materials should be addressed to W.W. (email: welyman@iphy.ac.cn), J.L. (email: jlluo@iphy.ac.cn) and G.L. (email:gtliu@iphy.ac.cn)

**Abstract**

We report the discovery of superconductivity on high-quality single crystals of transition-metal pnictides WP grown by chemical vapor transport (CVT) method. Bulk superconductivity is observed at $T_c$ = 0.84 K under ambient pressure by electrical resistivity and AC magnetic susceptibility measurements. The effects of magnetic field on the superconducting transitions are studied, leading to a large anisotropy parameter around 2 with the in-plane and out-of-plane upper critical fields of $H_{c2,\parallel}$=172 Oe and $H_{c2,\perp}$=85 Oe, respectively. Our finding demonstrates that WP is the first superconductor in 5d transition-metal at ambient pressure in MnP-type, which will help to search for new superconductors in transition-metal pnictides.

## Introduction

The novel and intriguing behaviors in transition-metal pnictides have long attracted the attention of researchers due to multiple quantum orders and competing phenomena. One of the examples is the recently discovered pressure-induced superconductivity in CrAs and MnP [1-4]. At ambient conditions, both CrAs and MnP adopt the MnP-type (B31) structure, belonging to the large family of transition-metal pnictides with a general formula MX. Most of the MX compounds crystallize in either the hexagonal NiAs-type structure or the orthorhombic MnP-type structure, for which two the second structure is a distorted first structure with small shifts of atomic positions [5]. By analyzing the structures of these phosphides as a function of the electron number per formula unit, Roald Hoffmann [6] found that the MnP-type phases can only exist in the transition-metal phosphide compounds with the number of valence electrons ranging from 11 to 14, such as CrP, MnP, FeP, CoP, RuP, WP, and IrP. Among these phosphides, MnP and FeP are metals exhibiting a unique magnetic structure consisting of a helical arrangement of spins with net antiferromagnetic ordering [7-8]. CrP and CoP are normal metals with Fermi-liquid ground states [9-10]. RuP is a metal of 4d transitional-metal phosphides with charge density wave (CDW) order [11]. Compared to that of 3d transitional-metal phosphides, the relatively wide energy band and the extended wave function in 5d transition-metal pnictides, such as WP and IrP, lead to a smaller density of states (DOS) and overlap of 5d-orbital. Therefore, the electrons are itinerant and no magnetic ordering was observed in 5d transition-metal pnictides [12]. To date, superconductivity is found only existing in pressured-MnP and Rh-doped RuP among phosphide compounds [13]. Moreover, the similarity of the superconducting phase diagrams between CrAs/MnP and Fe/Cu-based superconductors points to possible unconventional superconductivity [14-15]. Therefore, it is of high interest to explore whether intrinsic superconductivity at ambient pressure can be achieved in the MnP family.

Here we report superconductivity observed in high-quality WP single crystals at ambient pressure in the MnP family for the first time. The electrical resistivity and AC magnetic susceptibility measurements revealed a bulk superconductivity with $T_c$=0.84 K. The magnetotransport measurements showed that upper critical fields have an anisotropy parameter of around 2. The small electron-phonon coupling constant $\lambda$=0.45 obtained from temperature dependence of resistivity indicates that WP is a weak coupling Bardeen-Copper-Schrieffer (BCS) superconductor.

**Experimental Details**

The polycrystalline samples were prepared using a conventional solid-state reaction method. Stoichiometric amounts of W and P powders were mixed thoroughly and pressed into pellets. The pellets were then sealed in an evacuated quartz tube. In consideration of red phosphorus sublimates at about ~416 ℃ and tungsten reacts with oxygen, sulphur, and nitrogen only higher than 400 ℃, the sealed ampoule was heated to 400 ℃ over 10 hours and kept for 20 hours, then heated to 1000 ℃ over 40 hours and kept for about 3 days at this temperature before cooling down to room temperature. WP crystallizes in an orthorhombic structure with space group *P*nma (No. 62), as illustrated in Fig. 1a. The crystal structure of the compound was verified by Rietveld refinement [16] of powder x-ray diffraction (XRD) pattern collected on polycrystalline WP at ambient conditions. As shown in Fig. 1b, all the diffraction peaks are fully explained using a single phase in *P*nma (No. 62) with lattice parameters *a*= 5.7222(6) Å, *b*=3.2434(9) Å, and *c*=6.2110(6) Å, which are in good agreement with previously reported values [17,18].

Due to the high melting point of WP, it is very difficult to grow WP single crystal using conventional melting technique, such as the Bridgman method. For WP single crystal growth, we employed chemical vapor transport (CVT) method which has been successfully used to grow pnictides such as CrP, MoP, NbP and TaP [19-21]. The well-mixed starting materials of Iodine and WP polycrystalline powders were placed at one end of a 20 cm-length quartz tube and sealed under vacuum. After this process, the furnace was naturally cooled to room temperature by turning off the power. Finally, shiny needle-like single crystals were found to collect at the source end. By this method, we successfully synthesize the large single crystals of WP with a typical dimension of 0.1×0.1×4.0 mm as shown in Fig. 1c. The *b*-axis direction relative to the crystal morphology was determined from a single crystal lying on a beveled silicon slice on a powder diffractometer. The XRD pattern in the inset of Fig. 1(b) shows only (*l*00) peaks with no impurities, indicating the high quality of WP single crystal and the *b*-axis is parallel to the needle-shaped sample.

The electrical resistivity was measured with the conventional four-probe method using a Quantum Design Physical Property Measurement System (PPMS) from 300 K to 2 K. For temperatures below 2 K, the experiments were performed on a top-loading He3 refrigerator with a superconducting magnet up to 15 T. AC magnetic susceptibility was measured by a mutual inductance coil. The sample and a similar volume piece of lead (Pb) were put into the coil and fixed by glue. The diamagnetic signal due to the superconductivity transition was estimated by comparing with the diamagnetic signal of Pb. An excitation current of 0.5 mA with a frequency of 1117 Hz was applied to the primary coil and output signal across two oppositely wound secondary coils was picked up with a Standford Research SR830 lock-in amplifier.

**Results and discussion**

Figure 2a shows the temperature dependence of electrical resistivity measured on three typical WP single crystals (S1, S2, and S3) between 300 and 0.3 K in zero field. The WP has a normal-state resistivity ~40 μΩ cm at room temperature with a residual resistivity ratio (RRR) defined as $R_{300K}/R_{1.5K}$ of ~37. A linear dependence of resistivity $\rho(T)$ on temperature has been observed in the temperature range from 300 K to 60 K. Then $\rho(T)$ further decreases to a temperature-independent constant as the temperature reduced to 1.5 K, indicating that WP has a metallic conduction property. As the sample is further cooled, $\rho(T)$ drops steeply at about 0.85 K, signaling the onset of superconductivity. From the upper left upper inset of Fig. 2a, it can be seen that the superconducting transitions for these 3 samples are very sharp. Especially for sample S3, the transition width is less than 20 mK, indicative of the high quality of the sample. If we define the superconducting transition temperature $T_c$ as the temperature at the midpoint of the resistive transition, it is found that $T_c$ for these samples are very close ranging from 0.834 K and 0.840 K, further demonstrating the samples are uniform and of high quality.

The superconducting transition in WP single crystal is further characterized by the low-temperature AC susceptibility measurement. Figure 2b shows the magnetic field dependence of AC susceptibility $\chi$. As the external magnetic field $B$ decreases to 1000 Oe, a broad diamagnetic signal corresponding to the superconductivity of Pb is observed. Then when the magnetic field approaches to $B$=110 Oe, another large diamagnetic signal was detected, which is ascribed to the superconductivity of WP. Moreover, the field is in good agreement with the upper critical field measurement which will be shown in the following section. Compared to Pb, the signal of $\chi$ from WP is much stronger and sharper, further demonstrating the high quality of the WP single crystal and bulk superconductivity in nature.

To explore the superconducting phase diagram of WP single crystal, the temperature-dependent resistance $R(T)$ is systematically measured under different external applied magnetic fields. The inset of Fig. 3a shows the corresponding results for sample S3 under the magnetic field from 0 to 125 Oe. As the magnetic field increases, the superconducting transition temperature $T_c$ shifts to lower temperature and the superconductivity is completely suppressed at $B$=125 Oe. Figure 3a summarizes the upper critical field $\mu_0 H_{c2}$ as a function of temperature $T_c$, where $T_c$ is defined as the temperature at the midpoint of the resistive transition. A linear temperature dependence is observed for $H_{c2}$, which can be well fitted by the Werthamer–Helfand–Hohenberg (WHH) equation given by:

$$\mu_0 H_{c2}(0) = -A T_c \frac{d\mu_0 H_{c2}}{dT}\Big|_{T=T_c}$$

where $A$ is −0.693. As shown by the dashed lines, the best fittings yield $H_{c2}(0)=$ 118.0, 90.4, and 124.9 Oe for sample S1, S2 and S3, respectively. A similar observation has been made in MnP [4]. The obtained $\mu_0H_{c2}(0)$ allows us to estimate the Ginzburg-Landau coherence length $\xi$ = 1642 Å according to the relationship: $\mu_0H_{c2}(0) = \Phi_0/2\pi\xi^2$, where $\Phi_0$=2.067 ×$10^{-15}$ Wb is the magnetic flux quantum.

The anisotropic upper critical field of WP single crystal is confirmed by the experiments with tilted magnetic field. The upper left inset of Fig. 3b shows the angular dependent $R$(H) at 0.3 K, where $\theta$ is the tilt angle between the normal of the sample plane and the direction of the applied magnetic field (Right lower inset of Fig. 3b). The superconducting transition shifts to higher field with the external magnetic field rotating from perpendicular ($\theta$=0°) to parallel ($\theta$=90°) direction. Figure 3b plots the upper critical field $\mu_0H_{c2}$ extracted from the inset of Fig. 3b as a function of the tilted angle $\theta$. Clearly, the out-of-plane upper critical field $\mu_0H_{c2,\perp}$ is much smaller than the in-plane upper critical field $\mu_0H_{c2,\parallel}$, demonstrating a strong anisotropic transport properties in WP single crystal.

To access the superconducting mechanism of WP, we performed fitting with the Bloch-Gruneisen function [22],

$$\rho(T) = \rho(0) + A(\frac{T}{\Theta_D})^5 \int_0^{\frac{\Theta_D}{T}} \frac{T^5}{(e^T-1)(1-e^{-T})} dT$$

where A is a constant and $\Theta_D$ is an effective Debye temperature. From the lower right inset of Fig. 2a, one can see that the experimental data can be well described by the Bloch-Gruneisen function. The best fit gives the Debye temperature $\Theta_D$= 238 K, from which the electron-phonon coupling strength $\lambda_{ep}$ can be estimated from the McMillan formula [23],

$$\lambda_{ep} = \frac{1.04 + \mu^*\ln(\frac{\Theta_D}{1.45T_c})}{(1-0.62\mu^*)\ln\left(\frac{\Theta_D}{1.45T_c}\right) - 1.04}$$

where $\mu^*$ is the repulsive screened Coulomb potential and is assigned a value in the range 0.1-0.15. Setting $\mu^*= 0.13$ and $T_c$=0.84 K, the $\lambda_{ep}$ is calculated to be 0.453, implying that WP is a weak-coupling Bardeen-Copper-Schrieffer (BCS) superconductor.

In order to understand the superconductivity observed in WP, it is necessary to compare the interatomic distances between the nearest cations and the orbital overlap in the MnP-type (B31) structure including CrAs, CrP, and WP. As illustrated in Fig. 5, the transitional metal atom has three nearest neighbor metal atoms. Apparently, the largest interatomic distance is the *b*-axis length, which determines the physical properties of the MnP-type compounds. J. B. Goodenough successfully explained the magnetic properties of the MnP-type (B31) structure by introducing the crystal-field and ligand field theory [24-25]. At ambient pressure, CrAs has a relatively large *b*-axis length of *b*=3.445Å that leads to a small 3d-orbital overlap between $\boldsymbol{d_{x^2-y^2}}$ bands, therefore, the 3d-electron are localized which accounts for the magnetic ordering in CrAs. However, as the *b*-axis length decreases to a critical value $b_c$=3.37 Å under a pressure of *P*=1.0 GPa [5], the larger orbital-overlap between $3\boldsymbol{d_{x^2-y^2}}$ make CrAs adopt a collective state of the electrons without magnetic ordering and superconductivity emerges. As *P* increases to 5.0 GPa, *b* is further reduced and the superconductivity disappears. The *b*-axis length of CrP is only 3.11 Å, which is equivalent to that of CrAs under *P*=15.0 GPa [27], and no superconductivity was observed. This indicates that the variation of the *b*-axis length plays an important role in determining the physical properties of MnP-type compounds. Compared to the pressured-driven superconducting CrAs and normal metal CrP, the intermediate *b*-axis length of *b*=3.25 Å in WP, as shown in Fig 5, suggests that superconductivity may emerge. On the other hand, according to Matthias rule [28], the superconductivity prefers to emerge at a peak density of states (DOS) at the Fermi level. However, CrP has a dip-like DOS [6,27], which contradict Matthias rule. Additionally, CrP is far away from spin fluctuation dome. Combined the above two facts, it can be understood that why CrP is not a superconductor. Compared to Pauli paramagnetism observed in CrP [5], WP exhibits diamagnetism in a broad range near room temperature [10], this implies that the band structure effect decrease the effective mass *m*\* and Landau diamagnetism dominates the magnetic susceptibility. In comparison with 3d electrons in CrAs, 5d electrons in WP have an extended wave function and larger orbital overlap indicated by the blue lobes, this may be associated with the superconductivity observed in WP.

**Conclusion**

In summary, high-quality WP single crystals were successfully prepared by chemical vapor transport method. Combined the electrical resistivity and magnetic susceptibility measurements, we demonstrated for the first time that WP is the first intrinsic superconductor with $T_c$ =0.84 K in MnP-type family. Additionally, WP is the first superconductor in 5d-transition metal phosphides to date. Therefore, our study offers a platform to study the superconducting mechanism and may shed light to search for new superconductors in MnP-type compounds.

**Acknowledgements**

This work has been supported by the National Basic Research Program of China from the MOST under the grant No. 2015CB921101 and 2016YFA0300600, by the NSFC under the grant No. 11874406.

**Figure and Figure Caption**

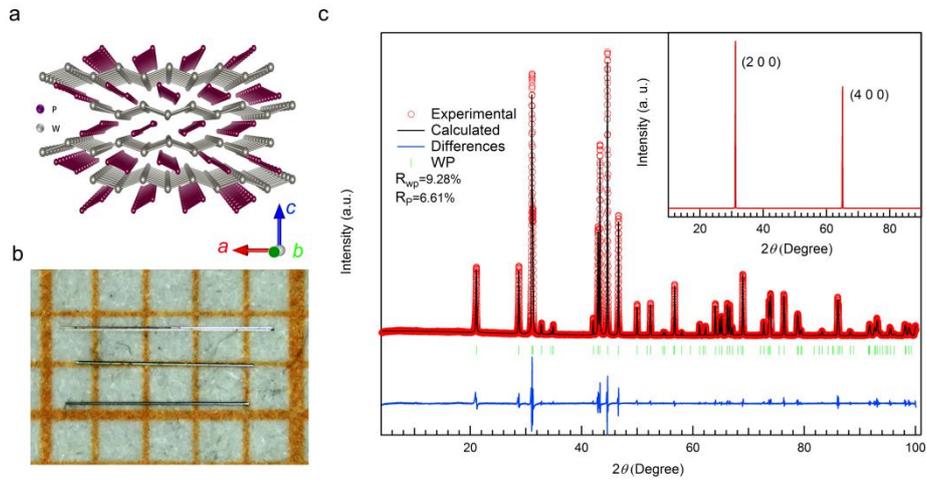

**Figure 1. Structural characterization of WP.** (a) Crystal structure of WP. (b) Room-temperature X-ray diffraction (XRD) spectra and the Rietveld refinement of polycrystalline WP. The open circle, solid line, and lower solid line represent experimental, calculated, and difference XRD patterns, respectively. The inset shows the XRD pattern for WP single crystals. (c) An optical image of the as-grown WP single crystals on a 1 mm × 1mm grid.

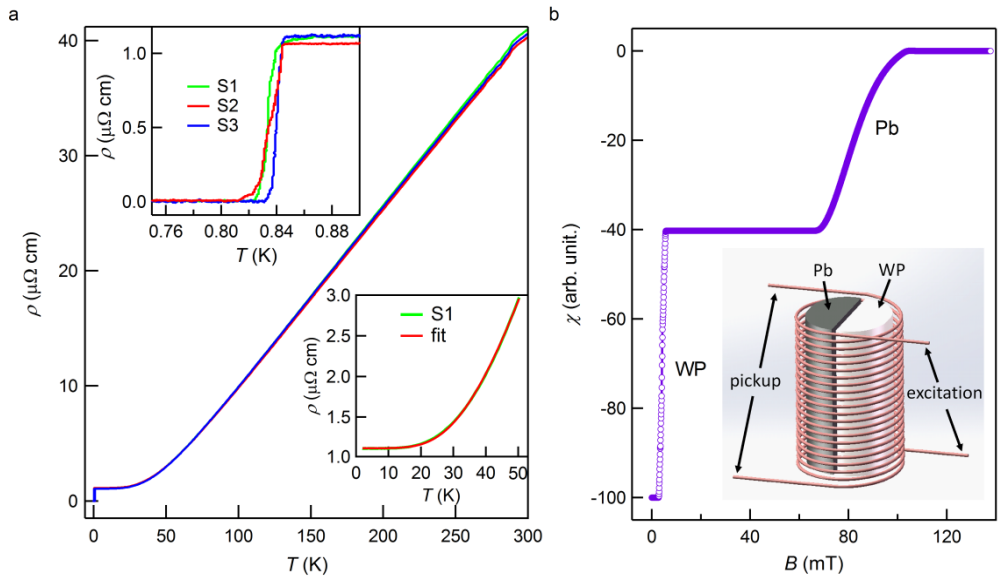

**Figure 2. Superconductivity characterization in WP single crystals.** (a) Temperature dependent resistivity $\rho$(T) for three typical WP single crystals (S1, S2, and S3) between 300 K and 0.3 K in zero external magnetic field. The upper left inset is an enlarged view of superconducting transitions at low temperatures. The lower right inset shows the comparison between the experimental data and the fitting result with the Bloch-Gruneisen function. (b) AC magnetic susceptibility for WP single crystal measured at 0.3 K. The inset is a schematic diagram of the AC magnetic susceptibility measurement setup.

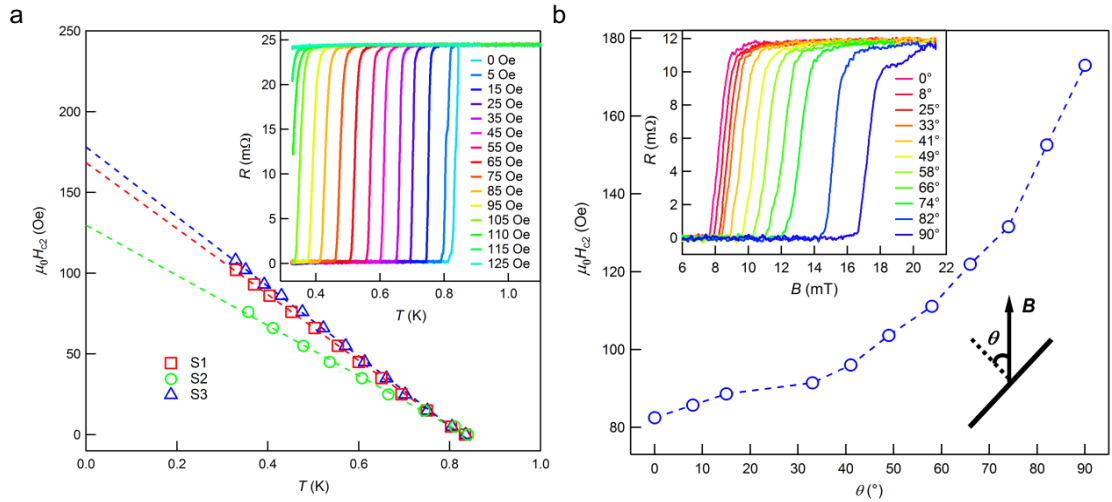

**Figure 3.** (a) Upper critical field $H_{c2}$ versus temperature phase diagram of WP single crystals. The dashed lines denote the Werthamer–Helfand–Hohenberg (WHH) fits to sample S1, S2, and S3, respectively. Inset: Temperature dependence of the resistance of sample S3 in different perpendicular magnetic fields. (b) Angular dependence of the upper critical field $\mu_0 H_{c2}$ for sample S1. The left upper inset shows the magnetic field dependence of the resistance for sample S1 measured at $T=0.3$ K with different tilt angle $\theta$. The right lower inset shows the scheme of our tilt experimental setup.

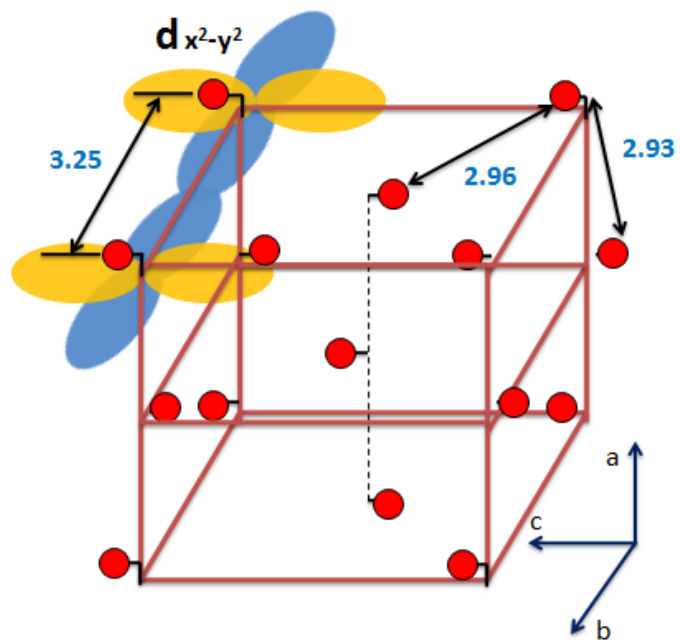

**Figure 5.** Cation sublattice of unit cell of WP with the indicated distances between the tungsten atoms. The components of $d_{x^2-y^2}$ along the $b$ and $c$-axis are represented by the blue and yellow lobes, respectively.